\documentclass[letter]{aa} % for the letters 
\usepackage{graphicx}
\usepackage{txfonts}
\usepackage{xcolor}
\newcommand{\kms}{km~s$^{-1}$}
\newcommand{\halpha}{H$\alpha$}
\newcommand{\lya}{Ly$\alpha$}
\newcommand{\HII}{\ion{H}{ii}}
\newcommand{\siioiii}{[\ion{S}{ii}]/[\ion{O}{iii}]}

\begin{document}

   \title{VLT/MUSE view of the highly ionized outflow cones in the nearby starburst ESO338-IG04\thanks{Based on observations collected at the European Southern Observatory at Paranal, Chile (ESO program 60.A-9314).}}

%   \subtitle{}

   \author{A. Bik
             \inst{1}
          \and
        G. \"{O}stlin
                \inst{1}
           \and
        M. Hayes
                \inst{1}
                \and
        A. Adamo
                \inst{1}
                \and
           J. Melinder
                \inst{1}
                \and
        P. Amram
                \inst{2}
          }

   \institute{Department of Astronomy, Oskar Klein Centre, Stockholm University, AlbaNova University Centre, SE-106 91 Stockholm, Sweden\\
              \email{arjan.bik@astro.su.se}
        \and
  Aix Marseille Universit\'{e}, CNRS,  LAM (Laboratoire d'Astrophysique de Marseille), F-13388, Marseille, France}

   \date{ }

% \abstract{}{}{}{}{} 
% 5 {} token are mandatory
 
  \abstract
  % context heading (optional)
   { The \lya\ line is an important diagnostic for star formation at high redshift, but interpreting its flux and line profile is difficult because of the resonance nature of \lya. Trends between the escape of \lya\ photons and dust and properties of
the interstellar medium (ISM) have been found, but detailed comparisons between \lya\ emission and the properties of the gas in local high-redshift analogs are vital for understanding the relation between \lya\ emission and galaxy properties.}  
   {For the first time, we can directly infer the properties of the ionized gas at the same location and similar spatial scales of the extended \lya\ halo around the local \lya\ emitter and Lyman-break galaxy analog ESO\,338-IG04.}
   {We obtained VLT/MUSE integral field spectra. 
   We used ionization parameter mapping of the \siioiii\ line ratio and the kinematics of \halpha\ to study the ionization state and kinematics of the ISM of ESO\,338-IG04.}
   {The velocity map reveals two outflows, one toward the north,
the other toward the south of ESO 338. The ionization parameter mapping shows that the entire central area of the galaxy is highly ionized by photons leaking from the \HII\ regions around the youngest  star clusters. Three highly ionized  cones have been identified, of which one is associated with  an outflow detected in the \halpha. We propose a scenario where the outflows are created by mechanical feedback of the older clusters, while the highly ionized gas is caused by the hard ionizing photons emitted by the youngest clusters. A comparison with the \lya\ map  shows that the (approximately bipolar)  asymmetries observed in the \lya\ emission are consistent with the base of the outflows detected in \halpha. No clear correlation with the ionization cones is found.} 
   {The mechanical and ionization feedback of star clusters significantly changes the state of the ISM by creating ionized cones and outflows. The comparison with \lya\ suggests that especially the outflows could facilitate the escape of \lya\ photons.}
 

   \maketitle
%
%________________________________________________________________

\section{Introduction}
 \lya\ is one of the most important diagnostic lines in extragalactic astrophysics. The line is thought to be bright in star-forming galaxies, allowing their detection at high redshift.  However, due to the resonance nature of the \lya\ line, relating the line strength to star formation is not a straightforward process, and  \lya\ needs to be used only with great care as a physical probe.
 A \lya\ photon scatters in neutral \ion{H}{I} gas until it escapes or is absorbed by dust. This means that  \lya\ is a complicated function of not only the  ionizing power and extinction, but also of the amount \citep{Pardy14} and kinematics  \citep{Kunth98,Wofford13} of \ion{H}{i}.
 
To understand the relation between \lya\ and these properties, detailed studies of local analogs of \lya -emitting galaxies are needed \citep{Ostlin14,Hayes14}. Observations of nearby galaxies have shown that stellar feedback creates highly ionized cones in the neutral medium, allowing the ionized radiation to escape from the galaxy \citep{Zastrow11,Zastrow13}.  This would make the amount of escaping Lyman-continuum (LyC) photons a strong function of the orientation under which we observe the galaxy. 

In this letter,  we compare  for the first time the spatially resolved \lya\ emission, obtained with the Hubble Telescope (HST) \citep{Hayes05,Ostlin09}, with the ionization state and kinematics of the ionized gas in the halo of the brightest \lya\ emitter in the local Universe and Lyman-break galaxy (LBG) analog, the starburst galaxy ESO\,338-IG04 (Tololo\,1924-416, hereafter ESO\,338). We present high-quality MUSE observations allowing a detailed comparison between the spatial distribution of the \lya\ emission and potential outflows  in this galaxy.  In Sect. 2 we describe the observations and data reduction. In Sect. 3 we present the results of the ionization mapping and kinematics of \halpha\ toward the galaxy. We conclude with  discussion and conclusions in Sect. 4.

  \begin{figure*}[!t]
   \centering
   \includegraphics[width=0.98\hsize]{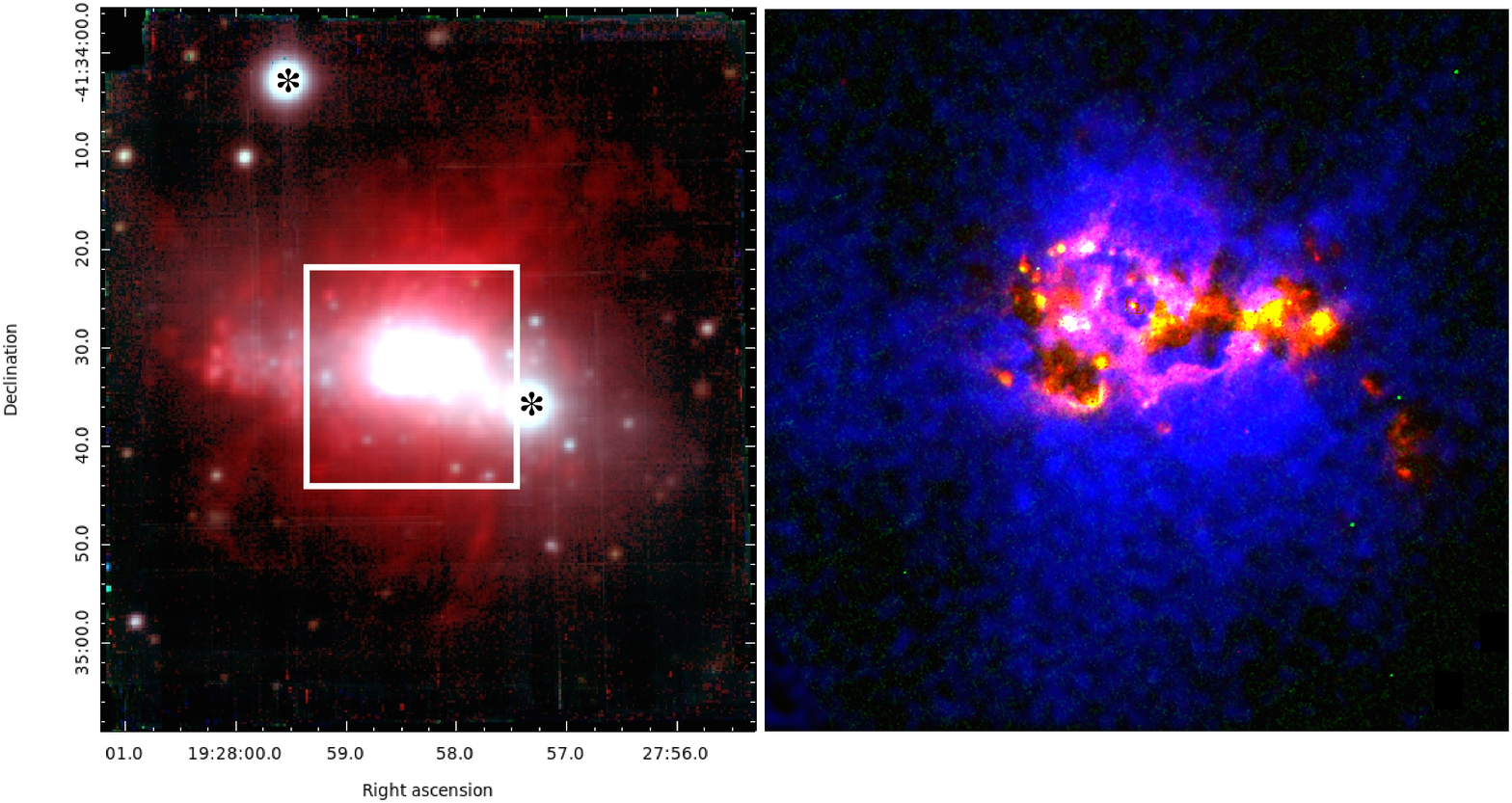}
      \caption{\emph{Left:} Reconstructed $VR$\halpha\ color image of ESO\,338 as seen by MUSE. The \halpha\ image is not  continuum subtracted.  The black asterisks mark the location of the two foreground stars. The white box shows the region displayed in the panel to the right. \emph{Right:} Reconstructed  three-color images based on narrow-band images taken with HST of \lya\ (blue), [\ion{O}{iii}] $\lambda$5007 \AA\ (green) and \halpha\ (red).}
         \label{fig:linemaps}
   \end{figure*}

   %__________________________________________________________________

\section{Observations and data reduction}

ESO\,338 has been observed with the integral field spectrograph Multi-Unit Spectroscopic Explorer (MUSE) \citep{Bacon10} mounted on the Very Large Telescope (VLT) on Paranal, Chile. The observations were performed during  the first science verification run. %(Prog Id. 60.A-9314). 
The observations are taken in the no-AO mode in the extended wavelength setting, providing spectra between 4600 and 9350 \AA\ in a field of view of 1 $\square$\arcmin\ with a 0.2\arcsec pixel scale.  The observations  were performed on 2014 June 25, split over two observing blocks with four frames of 750 sec, rotated by 90\degr, resulting in an  integration time of 100 min. The measured seeing was 0.9\arcsec in the $V$ band.
The data were reduced using the ESO pipeline version 0.18.5 (Weilbacher et al, in prep.). Flux calibration is  based on observations of the spectrophotometric standard star GD 153.
% Due to data acquisition at low instrument temperature, slit 6 of IFU 10 is vignetted and contains very low flux. 
%The ESO provided trace tables for this slit have been used to get a wavelength calibration. The slit has been removed from the science data. 
%The data reduction will be described in more detail in a forthcoming paper. 

We present the first results derived from this rich data set.  Broadband images in $V$ and $R$ were extracted using the $V$ and $R$ transmission curves. Emission line maps of [\ion{S}{II}] $\lambda$6717 \AA\ and $\lambda$6731 \AA, [\ion{O}{iii}] $\lambda$5007 \AA,\ and \halpha\ $\lambda$6563 \AA\ were extracted. The emission line maps were constructed by numerically integrating below the emission line, and the continuum was subtracted by averaging the continuum blue- and redward of the line. To enhance the low-surface brightness features, we made use of the weighted Voronoi tesselations binning algorithm by \citet{Diehl06}, which is a generalization of the algorithm developed by \citet{Cappellari03} . A minimum signal-to-noise ratio (S/N) of 20 has been chosen for the [\ion{S}{II}] $\lambda$6717 \AA\  line with a maximum cell area of 100 pixels (4 $\square$\arcsec). This Voronoi pattern was applied to all emission line maps. A  Gaussian profile was fitted to the \halpha\ line using the MPFIT \citep{Markwardt09} to calculate the velocity map.

\section{Results}

Figure \ref{fig:linemaps} (left) shows a $VR$\halpha\ three-color image extracted from the MUSE observations of ESO\,338. 
The reconstructed $V$ and $R$ images reveal the continuum emission of the galaxy and show numerous  star clusters \citep{Ostlin98}. The \halpha\ map shows a large ionised halo around the galaxy \citep{Bergvall02}, extending  as far as 6 kpc from the center of the galaxy \citep[assuming a distance of 37.5 Mpc,][]{Ostlin98}.

%We use emission lines of different ionization potentials to trace the ionization structure in the gas in and around the galaxy.
 The right panel of Fig. \ref{fig:linemaps} shows a composite of HST observations of \halpha, \ion{O}{III,} and \lya\ of the central part of ESO 338. The MUSE observations allow us to study the gas far out in the halo, while the HST observations give detailed information about the small scales \citep{Hayes05,Ostlin09}.

  \begin{figure*}[!t]
   \centering
   \includegraphics[width=\hsize]{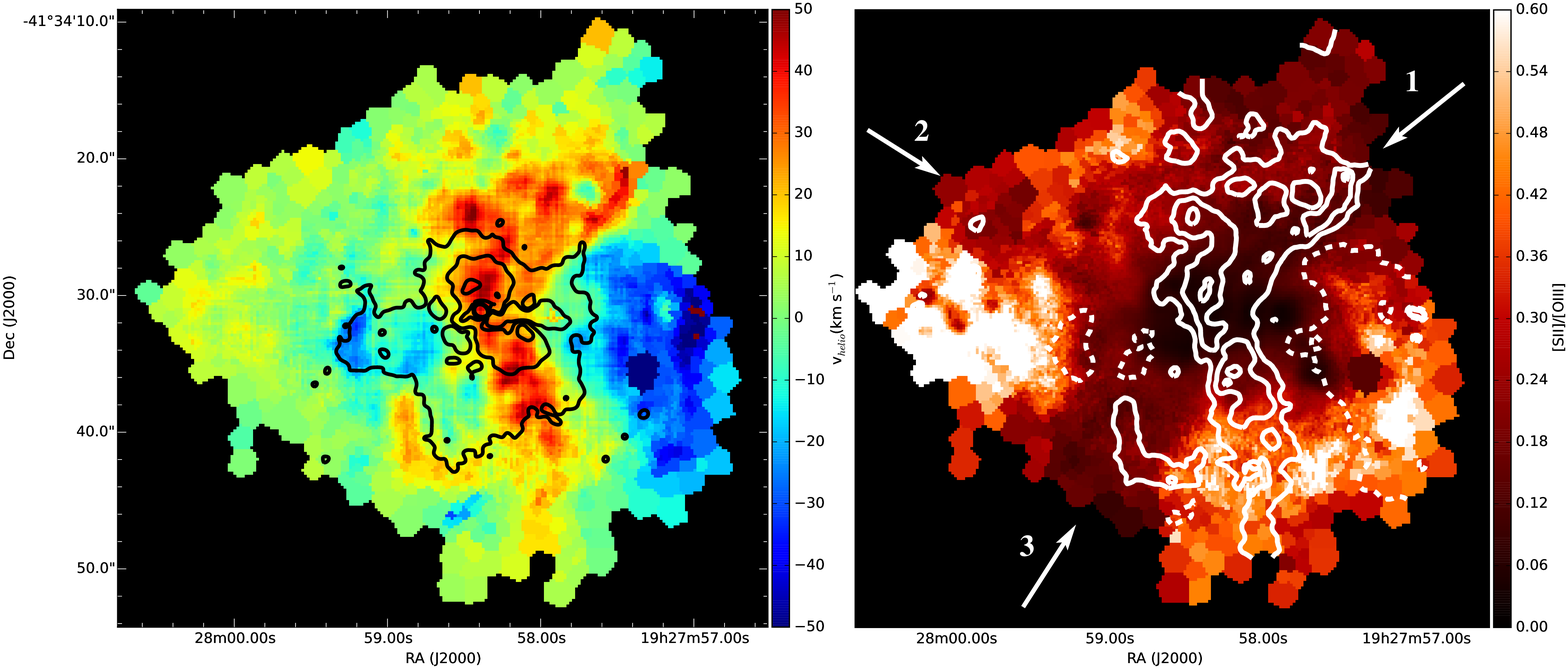}
      \caption{ \emph{Left:} Velocity map of \halpha\ as derived from a Gaussian fit of the emission line. As systemic velocity we have chosen 2841 \kms. Overplotted are the contours of the Ly$\alpha$ -continuum subtracted emission \citep{Ostlin09} with contour levels of 0.3, 0.9 and  $2\times10^{-19}$ erg/s/cm$^{2}$/\AA. \emph{Right:} \siioiii\ line ratio tracing the ionization parameter throughout the galaxy \citep{Pellegrini12}.   All cells with a flux lower than $1.7\times10^{-19}$ erg/s/cm$^{2}$/\AA, corresponding to a S/N = 20 for the \ion{S}{II} $\lambda$6717 \AA\ line, have been removed from the plot.  The contours show the \halpha\ velocity map with contours at -15 (dashed), 15, 30, and 45 (solid) \kms).
         \label{fig:ionization}}
   \end{figure*}

\subsection{Ionization parameter}

We applied the technique of ionization parameter mapping to determine the degree of ionization and the optical depth for Ly$\alpha$ photons using the  [SII] ($\lambda$6717 \AA+$\lambda$6731 \AA)/[OIII] ($\lambda$5007 \AA) line ratio \citep{Pellegrini12}. The right panel of Fig. \ref{fig:ionization} shows the \siioiii\ line ratio calculated using the Voronoi tessellated data.  The observed ratio varies from unity in the western and eastern parts of the halo to 0.03 in the center of ESO\,338. This extreme line ratio shows that there is a high production of $h\nu > $ 34 eV photons compared to where \siioiii\ is 1. Comparing these low values  with theoretical prediction by \citet{Pellegrini12} shows similarities to the models with an O3V star as ionizing star. The lowest values are found coincident with the positions of young ($<$2 Myr) star clusters \citep{Ostlin03}, containing these very early O stars.

The  central $\sim$2 kpc of the galaxy halo show an average value of 0.1 of the \siioiii\ ratio, indicating that entire center of ESO\,338 is highly ionized. This is probably caused by the \HII\ regions being optically thin for LyC photons, which allows them to leak out and ionize of the gas outside the \HII\ regions. The gas becomes optically thick for LyC photons only in the western and eastern outer regions of the galaxy. This pattern mimics the behavior of a giant \HII\ region, as found for similar galaxies \citep[Haro 11,][]{Cormier12}, and confirms the existence of an  in-out ionizing front that modifies the interstellar medium
(ISM) states of the galaxy. Several cones of highly ionized gas expanding all the way into the galaxy halo are visible. The most prominent ionization cones are in the northwest (nr\,1) and  southeast direction (nr\,3). An additional ionization cone (nr\,2) is found in the eastern direction. 

%Apart from the changes in ionization fraction of the gas, 
Changes for instance in extinction and abundance might also result in changes in line ratios. However, the extinction toward ESO\,338 is very low, with a luminosity-averaged value of close to 0 mag \citep{Ostlin03}. This is confirmed by our own H$\beta$/H$\alpha$ map, where maximum values of $E(B-V)$ = 0.2 mag are measured toward some clusters. To reproduce the observed change in \siioiii\ ratio, $E(B-V)$ values as high as 3.5 mag are needed \citep{Calzetti00}. To test whether the abundance variations might cause the observed pattern, we calculated the N$_{2}$ and S$_{3}$O$_{3}$  indices \citep{Stasinska06}. Spatial variations were found, but not in the same geometry as for \siioiii. 
%A detailed study of the abundances will be presented in a forthcoming paper.  
The same spatial variations, however, were found in line ratio maps of lines originating from different ionization potentials, such as \halpha/[\ion{O}{iii}] and [\ion{S}{ii}]/[\ion{S}{iIi}], suggesting that the  changes in the \siioiii\  ratio are dominated by variations in  ionization.

\subsection{Kinematics}

To relate the detected ionization cones to galactic-scale outflows, we compared the map of ionization parameters with the \halpha\  velocity map. Previous studies on the \halpha\ velocity field of ESO\,338 based on Fabry-Perot data \citep{Ostlin99,Ostlin01}  and long-slit spectroscopy \citep{Cumming08} 
%could not derive a clear rotation curve in the galaxy and 
concluded that either the galaxy is not in equilibrium, or the   \halpha\ velocities do not trace the gravitational potential and are dominated by feedback.

With our high-quality MUSE data we clearly detected several high-velocity outflows (Fig. \ref{fig:ionization}) and confirm that most of their observed velocity profile is indeed dominated by the outflow emission. 
%To derive velocities for the outflows, the systemic velocity of ESO\,338 needs to be determined. 
As a result of the irregular velocity profile, the systemic velocity is  only poorly defined. 
%Observations from neutral gas (21 cm, HIPASS) result in a systemic velocity of 2834 $\pm$ 5 \kms \citep[$z$ = 0.009453,][]{Meyer04}. 
We measured the velocity of the \halpha\ line of the intensity-weighted average over the part of the galaxy where continuum is visible.  This resulted in a observed velocity of 2841 $\pm$ 1 \kms, consistent with HIPASS 21 cm measurements \citep{Meyer04}.
Measurements of absorption lines of \ion{Na}{I} ($\lambda$ 5889.9\AA, 5895.9\AA) and \ion{Mg}{i} ($\lambda$ 5167.3\AA, 5172.7\AA, 5183.6\AA) in the intensity-weighted spectrum show similar values, albeit at lower accuracy because of their low equivalent width. This is consistent with previous studies where no velocity difference was found between stellar absorption lines and nebular emission  lines \citep{Cumming08,Sandberg13}.  After correcting for the heliocentric velocity, we chose the systemic velocity to be 2841 \kms\ as measured on the \halpha\ line of the integrated spectrum (Fig. \ref{fig:ionization}).% The velocity map shown in Fig \ref{fig:ionization} is calculated using this systemic velocity.

 A complex of outflows coincides with the northwestern ionization cone, suggesting that this cone is created by the expanding, highly ionized gas that is expelled from the galaxy by stellar feedback. A similar outflow is found to the south of the galaxy, but  this outflow is not aligned with an ionization cone. In addition, two expanding arcs are detected in the velocity map, one redshifted, roughly coinciding with ionization cone 3, and a blueshifted arc west of the same ionization cone.  The outflows show maximum projected velocities of 50 \kms, which is slower than outflows in M82 \citep{Shopbell98} and NGC\,1569 \citep{Westmoquette08},  for instance, which were derived from data with much higher spectral resolution. 
 %The MUSE spectral resolution is considerably lower, resulting in a average velocity of all the gas along the line of sight. 
 We found no evidence for  multiple components in  \halpha, probably because of the insufficient spectra resolution.

The MUSE data are much deeper than the previous spectroscopic \halpha\ observations of ESO\,338 of \citet{Ostlin99} and \citet{Cumming08}. This allowed us to trace the halo much farther out than before (6 kpc, 1\arcmin = 11 kpc). In addition to the outflows, we find evidence for a gradient in the velocity of the halo gas. The observed velocities  increase from  west to east between -50 \kms\ and +10 \kms, similar to the gradient observed in the \ion{H}{i} emission of ESO\,338 \citep{Cannon04}.

% This gradient could be explained by rotation of the halo gas around ESO\,338. 
%The observed velocities have been corrected to the $v_{lsr}$ rest frame (correction: 12.2 \kms). 

\section{Discussion and conclusions}

We presented MUSE integral field observations of the local \lya-emitting galaxy ESO\,338. We identified several highly ionized channels and found evidence of galaxy-scale outflows driven by stellar feedback in the central part of the galaxy.   ESO\,338 is perfectly suited for studying the relation  between the \lya\ escape and   ionization channels and outflows because it is the LBG analog closest to us. 

\subsection{Feedback}

A comparison between the ionization cones and the outflows shows that there is no one-to-one match. The northern outflow aligns with  ionization cone nr\,1.  The southern part of the galaxy halo shows a different geometry.  The outflow and ionization cone nr\,3 are not aligned and are probably two unrelated features. We did not detect velocity changes  for ionization cone nr\,2.  

The two arc features detected in the velocity map might be part of the same expanding structure. Figure \ref{fig:ionization} shows that the blueshifted  arc marks the edge of the highly ionized region, while the red arc is located at the back side of the galaxy expanding away from us. With this hypothesis, they could trace the shock front created by the expansion of the hot, overpressured gas from the center of the galaxy into the more neutral ISM.

Massive stars are sources of feedback, therefore they are the main agents of the modification of the ISM on both local and galactic scales. The large majority of massive stars are in clusters because ESO 338 has a very high cluster formation efficiency \citep[50\%,][]{Adamo11}. Hard ionizing photons are only produced by the most massive stars, which in turn means that they are
produced only in clusters with ages younger than $\sim$ 4 Myr, while the mechanical energy released by stellar winds and supernovae remains roughly constant for $\sim$ 30 Myr \citep{Leitherer99}.  \citet{Ostlin03} derived the star formation history based on the detected clusters and showed that the star formation increased 20-30 Myr ago and again 6-10 Myr ago, and it increased a little during the last 3 Myr. The oldest star formation burst has injected the most mechanical energy into the ISM and therefore might  be responsible for several of the observed outflows. The LyC photons responsible for the ionized cones are created by the most recent increase in  star formation. These channels might be outflow cones created by the older generation of  clusters and might be re-ionized by the younger generation. 

That not all ionization cones are associated with outflows might also be caused by  projection effects. 
The ionization cones are detected at any inclination because what we measure is the average ionization along the line of sight. Only if the  cone originates in the center and is directed straight toward us will the cone  not be detected because the center is already highly ionized.  The detection bias introduced by velocity measurements is different, such  that if the outflows  are in the plane of the sky, they will not be detected in the velocity map. A more detailed comparison with higher spatial resolution between the outflows and clusters is needed to test these hypotheses.

\subsection{\lya\ escape}

The \lya\ emission map  of ESO\,338 (Fig. \ref{fig:linemaps}) shows strong emission and absorption toward the young star clusters in the center of the galaxy \citep{Hayes05}. The \lya\ photons are produced in the \HII\ regions surrounding the star clusters. The low \siioiii\ ratio of the central area of the entire galaxy shows that these \HII\ regions are optically thin and leak LyC photons.

As noted by \citet{Hayes05}, the clusters in ESO338 are surrounded by a diffuse halo of \lya\, caused by the scattering of \lya\ photons in the neutral ISM. In addition to this scattering halo, two flux enhancements in the Lya map are observed as well: One at the northern side of the galaxy and one toward the southwest, and a relation with possible outflows has been proposed \citep{Hayes05,Ostlin09}. With the MUSE data we can relate these asymmetries to the outflows detected in the \halpha\ velocity map. The inner parts of the two detected outflows in the velocity map line up with the \lya\ asymmetries.  The expanding gas in the outflow causes the resonance wavelength to shift away from the \lya\ wavelength, which allows the \lya\ -photons to escape instead of being scattered.

The observed outflows as well as the geometry of the \lya\ emission shows that the determination of the \lya\ escape fraction will depend strongly on the ISM geometry of the galaxy. If we observe the galaxy through an outflow cone, the observed \lya\ flux may be much higher than for the way we observe ESO\,338. This allows testing theoretical models that predict that the variations of \lya\  are a function of the viewing angle of the ionization cones \citep{Laursen09,Behrens14}. 
 
We did not observe a  correlation with the ionization cones, apart from where the ionization cone aligns with the northern outflow.
This might suggest that outflows are more effective in enhancing the \lya\ escape fraction than ionization cones, which are more likely to facilitate the LyC continuum escape \citep[e.g.,][]{Zastrow13}. 
Interestingly, ESO\,338 is also a good candidate galaxy in which
to study the emission of
LyC radiation.  \citet{Leitet13} estimated a LyC escape fraction of
$\approx$16\% based upon the residual intensity in the usually saturated \ion{C}{ii} $\lambda=1036$\,\AA\ absorption line.

\begin{acknowledgements}
We thank the anonymous referee for useful comments that helped to improve the paper. We acknowledge the ESO science verification team for support and execution of the observations. M.H. and G.\"{O} received support from the Swedish research council (VR) and the Swedish National Space Board (SNSB).  \end{acknowledgements}

%-------------------------------------------------------------------

\bibliographystyle{aa}
\bibliography{MUSE_338}

\end{document}